\documentclass[aps,prl,superscriptaddress,twocolumn]{revtex4}

\usepackage{amsmath,amssymb,graphicx,xcolor,bm,soul}
\usepackage{comment}
\usepackage{upgreek}
\usepackage{xfrac}
\usepackage{braket}
\usepackage{epstopdf}
\usepackage{bbold}
\usepackage{url}

\definecolor{orange}{RGB}{235, 129, 0}
\usepackage[bookmarks=true,
            bookmarksnumbered=false,
            bookmarksopen=true,
            breaklinks=true,
            pdfborder={0 0 0},
            final=true,
            backref=none,
            colorlinks=true,
            linkcolor=blue,
            menucolor=blue,
            citecolor=blue,
            urlcolor=blue]{hyperref}

\begin{document}

\title{Quantum simulation and ground state preparation for the honeycomb Kitaev model}

\author{Tatiana A. Bespalova}
\affiliation{ITMO University, St. Petersburg, 197101, Russia}

\author{Oleksandr Kyriienko}
\email{o.kyriienko@exeter.ac.uk}
\affiliation{Department of Physics and Astronomy, University of Exeter, Stocker Road, Exeter EX4 4QL, United Kingdom}

\date{\today}

\begin{abstract}
We propose a quantum protocol that allows preparing a ground state (GS) of the honeycomb Kitaev model. Our approach efficiently uses underlying symmetries and techniques from topological error correction. It is based on the stabilization procedure, the developed centralizer ansatz, and utilizes the vortex basis description as the most advantageous for qubit-based simulations. We demonstrate the high fidelity preparation of spin liquid ground states for the original Kitaev model, getting the exact GS for $N=24$ spins using 230 two-qubit operations. We then extend the variational procedure to non-zero magnetic fields, studying observables and correlations that reveal the phase transition. Finally, we perform dynamical simulation, where the ground state preparation opens a route towards studies of strongly-correlated dynamics, and a potential quantum advantage.
\end{abstract}

\maketitle

\textit{Introduction.---}Strongly-correlated materials are described by models of interacting electrons and spins~\cite{Witczak-Krempa2014}, where an efficient classical description is inapplicable due to the sign problem and long-range entanglement \cite{Loh1990}. 
An area where strong correlations represent a formidable obstacle is magnetism~\cite{Savary2017}. When couplings between spins compete with each other, frustration leads to exotic phases of matter~\cite{Balents2010,Tomasello2019}. A striking example is a quantum spin liquid (QSL) phase~\cite{Savary2017,Takagi2019}, where strong quantum fluctuations persist even at low temperature $T$. QSL is a potential candidate for modelling high-$T_c$ superconductivity, and sheds light on unconventional magnetic materials~\cite{Anderson1987,Lee2007}. They include honeycomb iridates (Na$_2$IrO$_3$, Li$_2$IrO$_3$) \cite{Jackeli2009,Chaloupka2010,Rau2016}, 4d transition metal-based materials ($\alpha$-RuCl$_3$)~\cite{Kubota2015,Janssen2017,Gordon2019}, and herbertsmithite~\cite{Norman2016}. From the theoretical perspective QSL materials are often described by spin lattice models with bond-dependent Heisenberg coupling~\cite{Savary2017,Takagi2019}. The corresponding models for honeycomb lattices are Kitaev models~\cite{Kitaev2006,Trebst2017,Hermanns2018}. In many cases spin liquid physics cannot be accessed by efficient classical approaches, as this requires studying low energy behavior and calls for an exact diagonalization (ED)~\cite{Hickey2019,Lauchli2011,Wietek2018}. The density matrix renormalization group can push this boundary~\cite{Yan2011,Jiang2012} at the expense of a truncated wave function. Here, quantum simulation (QS) offers an alternative solution~\cite{Georgescu2014,Houck2012,Altman2021}, where strongly-correlated models can be studied on inherently quantum devices at a scale inaccessible classically \cite{Ebadi2021,Gulia2021}.

Quantum algorithms for materials and chemistry allow low-$T$ properties of a many-body Hamiltonian to be accessed using a ground state preparation protocol (GSP)~\cite{OxfordRev,ZapataRev,ElfvingRev}. They can use different principles. Some favor superior scaling, while leading to increased gate and qubit counts and targeting a fault-tolerant implementation~\cite{ZapataRev,Aspuru-Guzik2005,Dobsicek2007,OBrien2019,Berry2018,Babbush2019}. Others exploit quantum dynamics and overlap measurements to learn effectively the low-$T$ physics on mid-scale quantum simulators~\cite{Parrish2019,Kyriienko2020,Stair2020,Cotler2019,Bespalova2021,Seki2021}. For near-term operation, the leading approaches are based on variational principles with a hybrid quantum-classical loop as introduced in the variational quantum eigensolver (VQE)~\cite{Peruzzo2014,OMalley2016,Kandala2017}. This variational GSP relies on optimizing a parameterized quantum circuit (\emph{ansatz}) that prepares a minimal energy state at optimal parameters~\cite{CerezoRev,BhartiRev,Parrish2019a,Elfving2020,Yalouz2020,Peng2021}. This operation mode is favored experimentally~\cite{OMalley2016,Kandala2017,Arute2020,Ganzhorn2019,Kokail2019,IonQ2020}, as the circuit depth is typically reduced.

The efficiency of VQE crucially depends on the ansatz choice \cite{Woitzik2020,Patti2021,Funcke2021,OakRidge2021}. The performance limit is posed by the non-convex multiparameter optimization, where regions of vanishing gradients (\emph{barren plateaus}) hinder efficient optimization at large circuit depth \cite{McClean2018,Cerezo2021}. Correctly parametrized circuits can perform GSP in reasonable depth for cases where deep circuits with unsuitable structures fail. In chemistry, ansatze guided by the coupled cluster theory can offer a good convergence \cite{IonQ2020,Metcalf2020,Bauman2020,Sapova2021}, though at the expense of depth due to the fermion-qubit mappings. Hardware efficient ansatze (HEA) VQE use shallow circuits that are easy-to-run \cite{Kandala2017}, while missing the symmetries and rapidly approaching 2-designs~\cite{Harrow2009}. Various ansatz search techniques were proposed based on adaptive generator screening \cite{Grimsley2019,Tang2019,Claudino2020,Suchsland2021}, evolutionary and pruning approaches \cite{Chivilikhin2020,Zhang2021,Du2020,Sim2021,Bilkis2021,CChen2021,Ostaszewski2021}. Here, crucial advantage is offered by symmetry-preserving circuits~\cite{Gard2019,Barron2021,Herasymenko2019,Shkolnikov2021,HHSC2021}. For spin systems, many works target simpler models corresponding to transverse Ising, XXZ, and Heisenberg chains \cite{Kandala2017,WWHo2019,Grant2019,LasHeras2014,Kyriienko2018,Kardashin2021}. Recent developments comprise of the Hamiltonian variational ansatz for XXZ \cite{Wiersema2020}, natural gradient optimization for the Ising model \cite{Wierichs2020}, dynamical QS for 2D XY model \cite{Dumitrescu2021}, and tensor network ansatze for square Heisenberg \cite{Liu2019} and J$_1$-J$_2$ models \cite{Slattery2021}. Many more challenging spin models remain unexplored.

In this Letter, we propose a quantum protocol for studying the physics of quantum spin liquids. Developing an initialization technique and an ansatz motivated by symmetries, we simulate the ground state preparation at increasing scale, with an advantage of being verifiable. Utilizing digital QS, we study the dynamics of QSL in the strongly-correlated regime. Noting that developed techniques align with the roadmap towards quantum error-correction, we consider the simulation of spin liquids as a natural step towards advantage in material science.

\begin{figure}[t!]
\centering
\includegraphics[width=1.0\columnwidth]{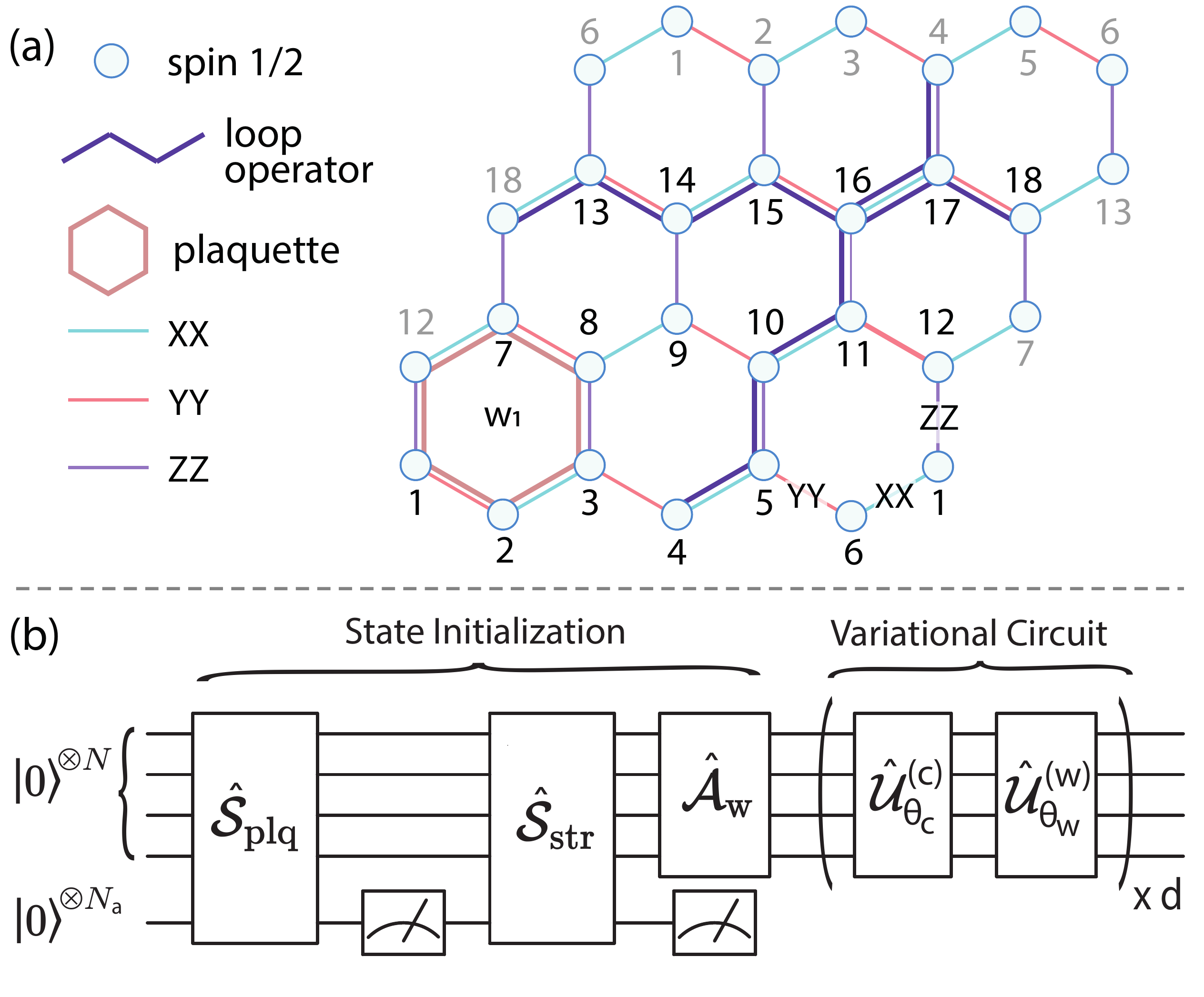}
\caption{(a) Kitaev lattice with $N=18$ spins arranged on a torus. Heisenberg couplings of ZZ, YY, and XX type act on bonds around hexagonal tiles (plaquettes). Plaquette operators (pink hexagons, $\hat{w}_p$) and loop operators (blue solid snakes) denote the conserved quantities. (b) Quantum protocol consisting of the stabilizer circuits $\hat{\mathcal{S}}_{\mathrm{plq}}$ and $\hat{\mathcal{S}}_{\mathrm{str}}$ that prepare an initial state with fixed number of vortices, which can modified by $\hat{\mathcal{A}}_\mathrm{w}$. Next, we use $d$-layers of variational centralizers ($\hat{\mathcal{U}}_{\bm{\theta}_c}^{\mathrm{(c)}}$) and vortex operations ($\hat{\mathcal{U}}_{\bm{\theta}_w}^{\mathrm{(w)}}$) that adjust the state to match the true ground state.}
\label{fig:sketch}
\end{figure}

\textit{The model.---}We study the Kitaev model of coupled spin-1/2s on a honeycomb lattice [see sketch in Fig.~\ref{fig:sketch}(a)]. Being exactly solvable, it represents a prototypical QSL example of quantum spin liquids~\cite{Savary2017}. However, in the presence of a local magnetic field the analytical solution is infeasible, and the system requires a numerical solution. The field-free honeycomb Kitaev Hamiltonian reads~\cite{Kitaev2006}
\begin{align}
\label{eq:H0}
\hat{\mathcal{H}}_0 = J^x \sum_{\langle i,j\rangle \in \mathcal{X}} \hat{X}_{i} \hat{X}_{j} +  J^y \sum_{\langle i,j\rangle \in \mathcal{Y}} \hat{Y}_{i} \hat{Y}_{j} + J^z \sum_{\langle i,j\rangle \in \mathcal{Z}} \hat{Z}_{i} \hat{Z}_{j},
\end{align}
where $\hat{X}_i$, $\hat{Y}_i$, $\hat{Z}_i$ are Pauli operators $\hat{P}_i^{\alpha}$ ($\alpha = x, y, z$), acting at site $i$ (or vertex or qubit). $J^{\alpha}$ are dimensionless interaction constants. $\mathcal{X}$, $\mathcal{Y}$, $\mathcal{Z}$ are sets of pairs of nearest neighbor sites (bonds), where XX, YY, and ZZ bonds alternate on a hexagon. They are shown in Fig.~\ref{fig:sketch}(a) by links of different color.
We consider the Kitaev model put on a torus, such that the lattice is periodic in both spatial dimensions. The lattice consists of $N$ vertices and $n = L_x \times L_y = N/2$ hexagonal \emph{plaquettes}, where $L_x$ ($L_y$) is the number of plaquettes in horizontal (vertical) direction. The toric boundary is arranged by adding two extra vertices, which for the $18$-qubit toric lattice are labeled as $17$ and $18$ [Fig.~\ref{fig:sketch}(a)]. 
The local magnetic field $\bm{h}$ with Cartesian components $h^{\alpha}$ is described by the bias terms
\begin{align}
\label{eq:Hm}
\hat{\mathcal{H}}_m = \sum_{i=1}^{N} h^x_i \hat{X}_{i} +  \sum_{i=1}^{N} h^y_i \hat{Y}_{i} + \sum_{i=1}^{N} h^z_i \hat{Z}_{i},
\end{align}
where the field may be inhomogeneous. The full Kitaev Hamiltonian then reads $\hat{\mathcal{H}} = \hat{\mathcal{H}}_0 + \hat{\mathcal{H}}_m$. In this work we concentrate on the isotropic Kitaev model $J^{x,y,z} = J$, while noting that our approach is also applicable to the anisotropic cases.
While the Kitaev model in the absence of magnetic fields described by $\hat{\mathcal{H}}_0$ is exactly solvable, its solution relies on describing the system in terms of Majorana fermions~\cite{Kitaev2006,Mandal2020}. Specifically, after performing the Jordan-Wigner transformation the Kitaev model can be reduced to free fermions in a static $\mathbb{Z}_2$ gauge field~\cite{ChenNussinov}. This was recently used for the simulation of braiding for Kitaev models in the momentum basis \cite{Kemper2020}. This mainly allows for predicting properties in the thermodynamic limit and calculating the spectrum~\cite{Lahtinen2008}, but does not provide a recipe for preparing the GS in the \emph{finite} system. Similarly to Hartree-Fock procedures in chemistry, free-fermion models have polynomial complexity, but if one wants to use their ground states as a stepping stone for correlated methods they do require a complex GSP~\cite{Arute2020}. Moreover, since mapping between qubits and fermions is non-local, this creates an overhead that inflates the QS budget of seemingly native spin-1/2 model. 
Instead, we show how the effective use of symmetries can yield a successful and scalable GSP protocol and open the route to dynamical simulations and QSL in non-zero field. 


\textit{Symmetries.---}The key insight for the efficient Kitaev model description comes from considering the vortex basis~\cite{Kells2008,Lahtinen2008,Lahtinen2011}. First, we can assign operators for each plaquette, where bond operators are multiplied counterclockwise along the chosen plaquette loop [e.g. see the plaquette operator $\hat{w}_1 = \hat{X}_1 \hat{Z}_2 \hat{Y}_3 \hat{X}_8 \hat{Z}_7 \hat{Y}_{12}$ in Fig.~\ref{fig:sketch}(a)]. 
The expectation value of the plaquette operator defines the presence or absence of a \emph{vortex} associated to each plaquette, as denoted by $-1$ and $+1$ expectations, respectively ($\hat{w}_p^2 = \mathbb{1}$). We note that plaquette operators commute with others, $[\hat{w}_{p}, \hat{w}_{p'}] = 0$ ($p, p' = 1,\dots,n$). They also commute with the Kitaev Hamiltonian, $[\hat{w}_p, \hat{\mathcal{H}}_0 ]=0~\forall~p$, and so does their sum, meaning that the number of vortices is conserved in the system (this symmetry is broken for $\hat{\mathcal{H}}_m$). We also have a constraint $\prod_p \hat{w}_p = \mathbb{1}$ such that vortices appear in pairs. The Hilbert space is thus fractured into separate manifolds labeled by the total number of vortices $W_{\mathrm{tot}}$, that is described by the expected value of the operator $\hat{\mathcal{W}}_{\mathrm{tot}} = ( n \mathbb{1} - \sum_{p=1}^{n} \hat{w}_p )/2$. 

Other two integrals of motion arise as \emph{loop} operators $[\hat{\ell}_{x,y}, \hat{\mathcal{H}}_0] = 0$ being the products of bond operators along two closed loops $\mathcal{L}_{x,y}$ on the torus. For example, we show the loop operators in Fig.~\ref{fig:sketch}(a) as solid blue lines corresponding to $\hat{\ell}_x = \prod_{i \in \mathcal{L}_x} \hat{Z}_i$ and $\hat{\ell}_y = \prod_{i \in \mathcal{L}_y} \hat{Y}_i$ strings~\cite{LeeFlammia2017}. The honeycomb Kitaev model can be seen as a topological subsystem code~\cite{Bombin2010}, where no logical qubits are encoded and loop operators are responsible for gauge fixing~\cite{Suchara2011,LeeFlammia2017}. We can define the Abelian stabilizer group $\mathcal{S} = \{\hat{w}_{p}\}_{p=1}^n \cup \{ \hat{\ell}_x, \hat{\ell}_y \}$ where all elements $\hat{S}_j \in \mathcal{S}$ pairwise commute. The physical state space of the Kitaev model corresponds to the codespace of $\mathcal{S}$ formed by eigenstates $\hat{S}_j |\psi\rangle = \pm |\psi\rangle$. Our goal is to utilize the symmetries when performing GSP.  


\textit{Variational search.---}To reach a ground state $|\psi_{\mathrm{GS}}\rangle$ of a strongly-correlated Hamiltonian we need to design a unitary $\hat{\mathcal{U}}_{\mathrm{GS}}$ that when acting on some initial state $|\psi_0\rangle$ prepares $\hat{\mathcal{U}}_{\mathrm{GS}} |\psi_0\rangle = |\psi_{\mathrm{GS}}\rangle$. For this we use a variational quantum protocol. During the variational search we train an ansatz circuit $\hat{\mathcal{U}}_{\bm{\theta}}$ parametrized by a vector $\bm{\theta}$ such that at optimal parameters $\bm{\theta}_{\mathrm{opt}} = \mathrm{argmin}_{\bm{\theta}} \langle \psi_{0}| \hat{\mathcal{U}}_{\bm{\theta}}^{\dagger} \hat{\mathcal{H}} \hat{\mathcal{U}}_{\bm{\theta}} |\psi_{0}\rangle$ the energy of the system is minimized, and $\hat{\mathcal{U}}_{\bm{\theta}_{\mathrm{opt}}} \approx \hat{\mathcal{U}}_{\mathrm{GS}}$. In the following, we compose an ansatz with $d$ layers of unitaries $\hat{\mathcal{U}}_{\bm{\theta}} = \prod_{k=1}^{d} \hat{\mathcal{U}}^{(w)}_{\bm{\theta}_w} \hat{\mathcal{U}}^{\mathrm{(c)}}_{\bm{\theta}_c}$. Here, $\hat{\mathcal{U}}^{\mathrm{(c)}}_{\bm{\theta}_c}$ is a unitary operator that preserves symmetries and $W_{\mathrm{tot}}$, essential property for fast GSP in the absence of the magnetic field. The unitary operator $\hat{\mathcal{U}}^{(w)}_{\bm{\theta}_w}$ introduces processes that do not conserve vortices, and is added when $h \neq 0$. 


\textit{State preparation.---}For an efficient variational approach we need to start in a suitable vortex number eigenspace. In the thermodynamic limit the set of degenerate ground states of $\hat{\mathcal{H}}_0$ corresponds to the zero-vortex sector $(W_{\mathrm{tot}} = 0)$, and this can differ for small systems.
To initialize the system we prepare a lattice of qubits as one of the eigenstates of the stabilizer group $\mathcal{S}$. The procedure originates from techniques used in topological quantum computing, showing that for topological subsystem codes based on the Kitaev model the stabilization can be performed by two-qubit projections~\cite{Suchara2011}. This resembles GHZ state preparation in the simple 2-qubit case, and 4-qubit stabilization for a rectangular grid of the surface code \cite{Fowler2012}. To apply projectors we use an ancillary qubit register with $N_a$ qubits. The initialization circuit is shown in Fig.~\ref{fig:sketch}(b), where unitary operators $\hat{\mathcal{S}}_{\mathrm{plq}}$ and $\hat{\mathcal{S}}_{\mathrm{str}}$, composed of CNOTs and Hadamards, are followed by ancilla measurements. After stabilization we can additionally apply $\hat{\mathcal{A}}_{\mathrm{w}}$ based on single-qubit rotations that enable pairwise creation and annihilation of vortices, adjusting $W_{\mathrm{tot}}$ to zero (or any other sector), and prepare $|\psi_0\rangle$ (see details in SM, sec.\,A).
\begin{figure}[t!]
\centering
\includegraphics[width=1.0\linewidth]{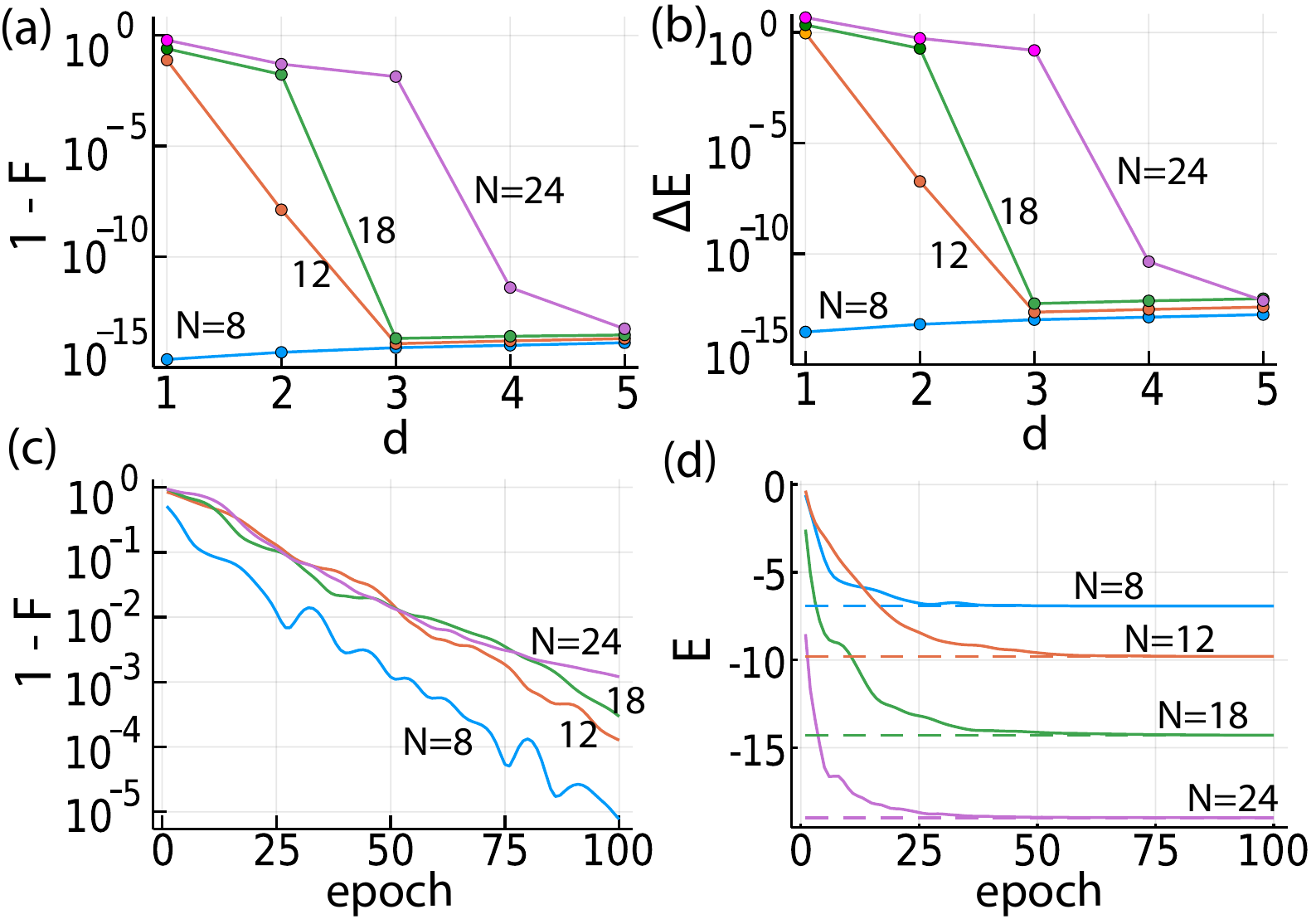}
\caption{Infidelity (a) and energy distance (b) between the variational and true Kitaev model GS, shown as a function of $d$ for the final epoch out of $500$. 
We consider $J=-1$, $h=0$, and four system sizes $N = \{8, 12, 18, 24 \}$. (c) Infidelity at different epochs and $N$. $d$ is chosen as the closest integer to $N/6$. (d) Energy vs epoch for the same parameters as in (c).}
\label{fig:nomagnetic}
\end{figure}


\textit{Ansatz construction.---}To remain in the relevant subspace we need to compose an ansatz circuit using generators that obey the symmetries of the Kitaev model, yet do not commute with $\hat{\mathcal{H}}_0$ and thus can change the energy. Such group operators are known as \emph{centralizers} $\mathcal{C}(\mathcal{S})$ of the stabilizer group $\mathcal{S}$. Namely, centralizers $\hat{C}_j \in \mathcal{C}(\mathcal{S})$ are bounded Pauli strings that commute with all the elements of $\mathcal{S}$. For the honeycomb Kitaev lattice these are bond operators $\hat{K}^{\alpha}_{(i,j) \in \mathcal{A}} = \hat P^{\alpha}_i \hat P^{\alpha}_j$ ($\alpha = x, y, z$ and $\mathcal{A} = \mathcal{X}, \mathcal{Y}, \mathcal{Z}$, correspondingly), such that $\big[\hat{K}^{\alpha}_{(i,j) \in \mathcal{A}}, \hat{w}_p \big] = 0$ for all $(i,j)$ and $p$. In fact, $\hat{K}^{\alpha}_{i,j}$ form only one possible subgroup of centralizers, and we can obtain others by multiplication of $\hat{K}^{\alpha}_{i,j}$. 
For $h = 0$ the ansatz is then composed of $d$ layers of centralizer unitaries $\hat{\mathcal{U}}^{\mathrm{(c)}}_{\bm{\theta}_c} = \hat{U}^{x}_{\bm{\theta}_{k,x}} \hat{U}^{y}_{\bm{\theta}_{k,y}} \hat{U}^{z}_{\bm{\theta}_{k,z}}$, where $\hat{U}^{\alpha}_{\bm{\theta}_{k,\alpha}} = \prod_{(i,j) \in \mathcal{A}} \exp(-i \theta_{k,\alpha,(i,j)} \hat{K}^{\alpha}_{i,j}/2)$. We consider that each variational angle is controlled independently, but note that correlated strategies can also be employed.

For the non-zero magnetic field the number of vortices is not conserved, and the GS of $\hat{\mathcal{H}}$ possesses a fractional number of vortices. Yet, it is convenient to use the vortex basis such that additional variational operations $\hat{\mathcal{U}}^{(w)}_{\bm{\theta}_w}$ are introduced as vortex creation and annihilation, as well as conditional vortex NOT (see SM, sec.\,B). As we show below, this ansatz is able to capture the physics of interacting vortices~\cite{Lahtinen2010,Lahtinen2011,Lahtinen2012} even for $h \sim J$.


\textit{Results: $h=0$.---}First, we consider a pure Kitaev model with zero magnetic field. We perform variational GSP using the proposed workflow, and compare the prepared state $|\psi_{\bm{\theta}} \rangle = \hat{\mathcal{U}}_{\bm{\theta}}|\psi_0\rangle$ with $|\psi_{\mathrm{GS}}\rangle$ obtained from imaginary time propagation~\cite{NoteKrylov}. We consider four different lattice sizes with $N = \{8, 12, 18, 24\}$ qubits, and prepare initial states $\{|\psi_0\rangle \}_N$ with $\langle\psi_0| \hat{\mathcal{W}}_{\mathrm{tot}} |\psi_0\rangle = \{4, 0, 0, 0 \}$, respectively, such that the ground state can be reached by the centralizer-based ansatz. We use a fast {\sffamily{}Julia}-based full state simulator realized in {\sffamily{}Yao}~\cite{Luo2019,Liu2019,Zeng2019,Kyriienko2021} and GPU acceleration for the GSP of a $24$-qubit lattice. The variational search is performed using the gradient-based optimizer Adam.
\begin{figure}[t!]
\centering
\includegraphics[width=1.0\linewidth]{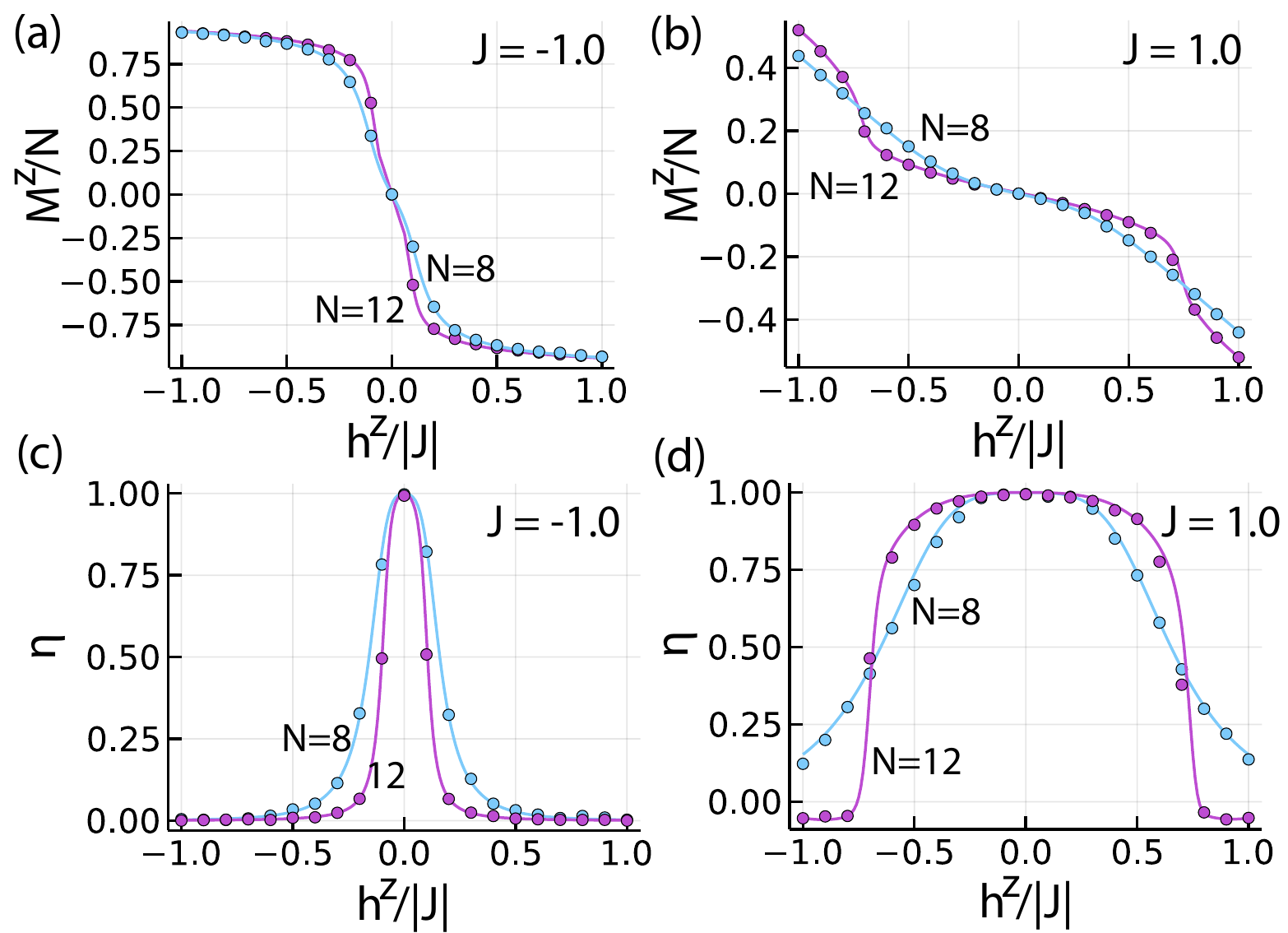}
\caption{(a, b) Normalized magnetization $M^z/N$ for the GS at uniform magnetic field $h^z$, shown for FM (a) and AFM (b) Kitaev model. Blue and purple dots correspond to $N=8$ and $N=12$ variational GSP with $d=50$ and $200$ epochs. Solid curves are for the true GS of $\hat{\mathcal{H}}$ obtained by imaginary time propagation. (c, d) Deviation of the number of vortices as a function of $h^z$, plotted for the same parameters as in (a, b).}
\label{fig:withmagnetic}
\end{figure}
The results are shown in Fig.~\ref{fig:nomagnetic}. As quality metrics we use an infidelity $1 - F = 1 - |\langle \psi_{\mathrm{GS}}|\psi_{\bm{\theta}}\rangle|^2$ and a GS energy distance $\Delta E$. In Fig.~\ref{fig:nomagnetic}(a) the infidelity is shown at $\bm{\theta}_{\mathrm{opt}}$ as a function of $d$ for different $N$. The proposed GSP shows an excellent performance, converging to the exact ground state (up to numerical precision) already at $d \sim N/6$. In Fig.~\ref{fig:nomagnetic}(b) we present the energy distance, showing perfect correspondence with the infidelity. Plotting the training as infidelity vs epoch number [Fig.~\ref{fig:nomagnetic}(c)], we observe exponential convergence and high-quality GSP even for the largest $N=24$ lattice, noting the absence of barren plateaus. Comparing results with HEA, we see that the latter struggles to prepare an approximate GS even for the minimal Kitaev lattice (see SM, sec.\,C). The convergence to relevant GS energies $E$ is shown in Fig.~\ref{fig:nomagnetic}(d).

\textit{Results: $h \neq 0$.---}Next, we prepare the Kitaev model GS with a non-zero uniform magnetic field, choosing $h^z \neq 0$. The magnetic field lifts the GS degeneracy and leads to quantum phase transitions between QSL and a trivial polarized phase~\cite{Hickey2019}. Due to the broken vortex conservation symmetry GSP becomes challenging, mirroring the complexity of classical spectrum calculations. By including vortex rotations and vortex-CNOTs in the ansatz, we prepare approximate ground states of $N = 8$ and $N = 12$ Kitaev lattices with $10^{-3}$ infidelity and comparable energy distance (see SM, sec.\,D). Performing GSP for increasing $h^z$ and both ferromagnetic (FM, $J = -1$) and antiferromagnetic (AFM, $J = 1$) Kitaev couplings, we show relevant system observables in Fig.~\ref{fig:withmagnetic}. Specifically, in Fig.~\ref{fig:withmagnetic}(a,b) we show the normalized total magnetization $M^z/N = \sum_{j=1}^{N} \langle \psi_{\bm{\theta}_{\mathrm{opt}}} | \hat{Z}_j | \psi_{\bm{\theta}_{\mathrm{opt}}}\rangle / N$ (dots), closely following true GS magnetization shown by solid curves. In Fig.~\ref{fig:withmagnetic}(c,d) we plot the average vortex number deviation $\eta = | 2 W_{\mathrm{tot}}/n - 1 |$, showing the transition between QSL physics to trivial physics (see SM, sec.\,D). Intriguingly, already at small system sizes we observe qualitative changes that resemble transitions in Ref.~\cite{Hickey2019}.
\begin{figure}[t!]
\centering
\includegraphics[width=1.0\linewidth]{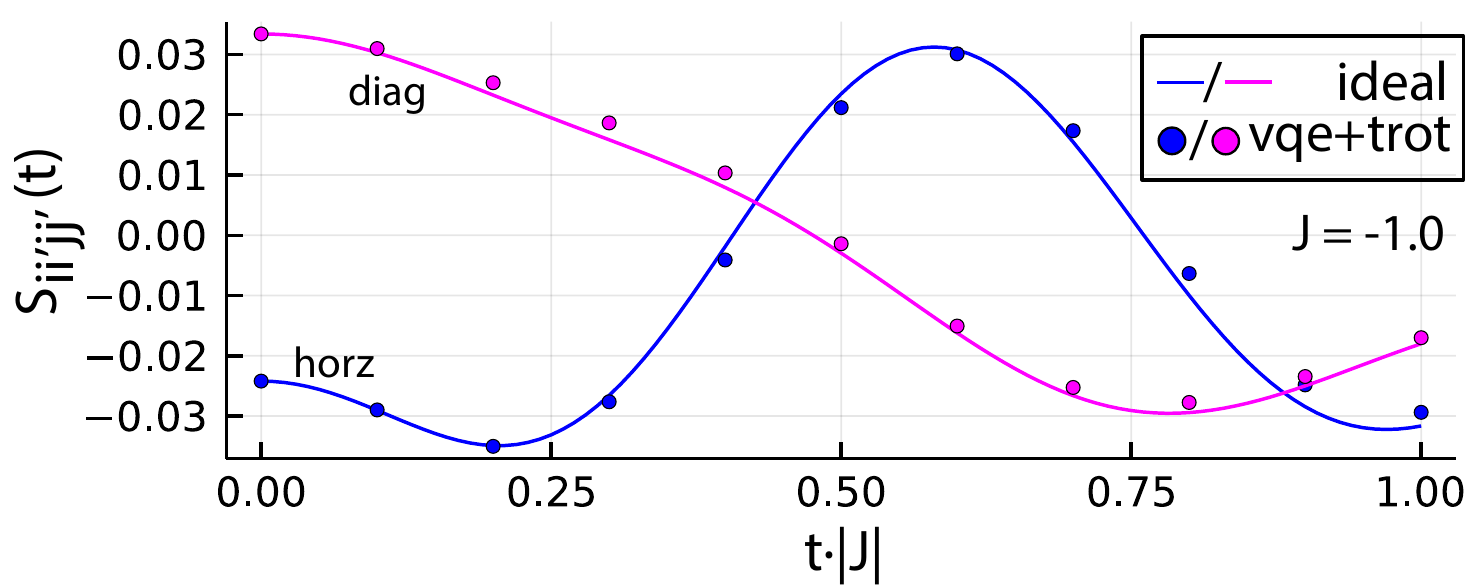}
\caption{Dynamical bond-bond correlation shown as a function of time for $N=12$ FM Kitaev model at $h^z=0.5|J|$. Magenta and blue curves show exact propagation from true GS with horizontal ({\sffamily horz}) and diagonal ({\sffamily diag}) bond orientations. The dots show variational GS ($d=3$ and $300$ epochs) propagated by Trotterization with steps of $0.1 |J|^{-1}$ duration.}
\label{fig:correlations}
\end{figure}

\textit{Results: dynamics.---}Finally, we generalize our considerations to study dynamical effects. Quantum simulators excel in evolving quantum systems, where a propagator $\hat{U}(t) = \exp(-i t \hat{\mathcal{H}})$ is approximated by a sequence of gates~\cite{Georgescu2014}. Then, preparing the Kitaev GS at zero field, simulation of dynamics for $h\neq 0$ opens the route to quench studies of strongly correlated spins~\cite{Hermanns2018}. Let us define bond-bond correlations functions~\cite{Yang2008,Baez2020} using time-dependent expectations of two-body operators $m_{kk'}(t) = \langle \psi_{\bm{\theta}}| \hat{U}^{\dagger}(t) \hat{Z}_k \hat{Z}_{k'} \hat{U}(t) |\psi_{\bm{\theta}}\rangle$ and four-body operators $c_{kk'll'}(t) = \langle \psi_{\bm{\theta}}| \hat{U}^{\dagger}(t) \hat{Z}_k \hat{Z}_{k'} \hat{U}(t) \hat{Z}_l \hat{Z}_{l'} |\psi_{\bm{\theta}}\rangle$. We define the static correlator as $C_{ii'jj'} = c_{ii'jj'}(0) - m_{ii'}(0) m_{jj'}(0)$, and the dynamics correlator as $S_{ii'jj'}(t) = c_{ii'jj'}(t) - m_{ii'}(t) m_{jj'}(0)$. Such correlations can be accessed by dynamical QS with a proven advantage over classical computation~\cite{Baez2020}.

In Fig.~\ref{fig:correlations} we show \emph{dynamical} bond-bond correlations for the variationally prepared GS of $\hat{\mathcal{H}}_0$ that is evolved under the biased Kitaev Hamiltonian with $h^z/|J| = 0.5$, simulating the quench. We propagate the system using the Trotterization approach with $10$ steps. The result is shown by blue and magenta dots for horizontal and diagonal bond orientations (SM, sec.\,E). We confirm that $S_{ii'jj'}$ can be successfully reproduced with the proposed workflow, following the solid curves obtained for the true GS and idealized QS. The results can be further improved if analog QS is used \cite{Schmied2011,Weimer2011}. Additionally, we studied \emph{static} bond correlations at $h \neq 0$ (SM, sec.\,D), and note that our results can be extended to track long-range spin-spin correlations~\cite{Baskaran2007,Mandal2011}. Also, similar protocol can be used to study QSL quasiparticles in non-zero fields~\cite{Knolle2014}.


\textit{Discussion and conclusion.---}We observe that for limited circuit depth we \emph{can} prepare high-quality QSL ground states in the noiseless limit. This is the first step towards scalable GSP and QS for Kitaev models. For this scaling to persist when implemented on a real hardware, we require a noise level that does not impede optimization. Given large gradients, we expect the procedure to work at relatively low shot numbers. We assume a gate set based on  rotations plus CNOTs, and assign a single-qubit and CNOT errors as $\varepsilon_1$ and $\varepsilon_2$. Using the simplest noise model, which was shown to describe large scale quantum experiments \cite{Arute2019}, we estimate fidelity as $F = (1- \varepsilon_1)^{n_{\mathrm{R}}} (1- \varepsilon_2)^{n_{\mathrm{CNOT}}}$, where $n_\mathrm{R}$ and $n_{\mathrm{CNOT}}$ is the number of rotations and CNOTs, respectively (see details in SM, sec.\,F).
For $N=18$, $\varepsilon_1 = 10^{-4}$ and $\varepsilon_2 = 10^{-3}$ we can achieve GS with $F = 82 \%$ fidelity, and for $\varepsilon_1 = 10^{-5}$ and $\varepsilon_2 = 10^{-4}$ this can improve to $F = 98\%$. Furthermore, we note that the coupling to environment may make QS more rich, and lead to exceptional spin liquids \cite{Bergholtz2021}. In SM, sec.\,F, we also consider the question of connectivity and boundary conditions, presenting possible mappings between the hexagonal lattice and hardware-native square~\cite{Arute2019} and heavy-hex~\cite{IBM2020} lattices. As for the initialization, we stress that studies of Kitaev's QSL physics perfectly aligns with efforts towards quantum error correction~\cite{Kelly2015,Corcoles2015,Bultink2020,Google2021,IonQ2020b}, where the stabilization procedure was demonstrated. Given that for zero magnetic field the model is classically tractable and can be verified at increasing $N$, can it become a benchmarking problem on the road to quantum advantage?

Recent state-of-the-art experiments include realizing topologically ordered toric code ground states on a quantum processor~\cite{Satzinger2021}, working at the scale of $31$ qubits. Here, we suggest Kitaev spin liquid as a material science challenge that may be tackled by modern quantum devices. \emph{In fine}, we proposed the efficient variational workflow for the ground state preparation utilizing symmetries of the Kitaev model, initialization, and vortex-based operations. When paired with dynamical quantum simulation, our toolbox allows studying the physics of strongly-correlated spins that is classically inaccessible.

\begin{acknowledgments}
\textit{Acknowledgements.---}We acknowledge valuable discussions with Laura Baez in the beginning of the project. We also thank Annie Paine for useful suggestions on the lattice mappings and reading the manuscript. T.\,A.\,B. thanks the University of Exeter for hosting her visit as a part of the MSc project work. 

\textit{Note added.---}While completing this work several independent variational protocols of kagome Heisenberg models and square-octagon-lattice Kitaev model appeared in \cite{Kattemolle2021}, \cite{Bosse2021}, and \cite{Li2021}. While having the same spirit and pushing NISQ boundaries, our work highlights the importance of stabilization and adds dynamical aspects.
\end{acknowledgments}

\end{document}